\documentstyle[12pt]{article}

\newcommand{\be}{\begin{equation}}
\newcommand{\ee}{\end{equation}}
\newcommand{\bn}{\begin{eqnarray}}
\newcommand{\en}{\end{eqnarray}}
\newcommand{\p}{\partial}
\newcommand{\pslash}{p\!\!\!/}
\newcommand{\nslash}{n\!\!\!/}
\newcommand{\aslash}{a\!\!\!/}

\begin{document}

\begin{center}

\noindent{\large\bf
{\huge The Jacobi identity for Dirac-like brackets}
}\vspace{6mm}

\noindent{
{\Large E. M. C. Abreu$^{}$, D. Dalmazi$^{}$ and
E. A. Silva$^{}$}
}\vspace{3mm}

\noindent

\begin{large}
{\it Departamento de F\'{\i}sica e Qu\'{\i}mica, UNESP/Guaratinguet\'a,  \\
Av. Ariberto Pereira da Cunha 333, Guaratinguet\'a, \\ 12516-410, S\~ao Paulo,
SP, Brazil, \\
{\sf E-mail: everton@feg.unesp.br and dalmazi@feg.unesp.br.}}\\
\end{large}
\vspace{4mm}

\today

\vspace{14mm}

\end{center}

\begin{abstract}
For redundant second-class constraints the Dirac brackets cannot be defined and new brackets must be introduced.  We prove here that the Jacobi identity for the new brackets must hold on the surface of the second-class constraints.  In order to illustrate our proof we work out explicitly the cases of a fractional spin particle in $2+1$ dimensions and the original Brink-Schwarz massless superparticle in $D=10$ dimensions in a Lorentz covariant constraints separation.
\end{abstract}

\newpage

\section{Introduction}

For covariance reasons we are often forced to work with more variables than the minimal set of physical fields.  The presence of spurious degrees of freedom are associated with local gauge  symmetries.  In the Hamiltonian formulation of gauge theories the gauge transformations are generated by the so-called first-class constraints according to Dirac's classification:
\be
\phi_a\,(q,p)\,\approx\,0 \qquad \qquad \qquad a\,=\,1,\ldots,k\;\;.
\ee
There are also second-class constraints (SCC) which are not associated with local symmetries but are responsible for the correct counting of degrees of freedom,
\be
\chi_{\alpha}\,(q,p)\,\approx\,0 \qquad \qquad \qquad \alpha\,=\,1,\ldots,n\;\;.
\ee
The first-class constraints have weakly vanishing Poisson brackets (PB) with the full set of $ (n+k)$-constraints, while the SCC have non-vanishing PB among themselves,
\be
\{\,\chi_{\alpha},\chi_{\beta}\}\,\equiv\,C_{\alpha\beta}\;\;.
\ee

Following Dirac we can eliminate the SCC by defining the now called Dirac brackets (DB):
\be
\{A,B\}_{DB}\,=\,\{A,B\}\;-\;\{A,\chi_{\alpha}\}\,(C^{-1})^{\alpha\beta}\,\{\chi_{\beta},B\}
\ee
the dynamics of the system can be entirely formulated in terms of DB.

By construction the SCC have vanishing DB with any quantity in the phase space: $\{A,\chi_{\alpha}\}_{DB}=0=\{\chi_{\alpha},B\}_{DB}$ and thus can be strongly set to zero.  The basic properties of the PB are inherited by the DB.  Perhaps the less trivial property is the Jacobi identity.  Using the definition of PB
\be
\{A,B\}\,=\,\omega_0^{IJ}\,\p_I A\, \p_J B
\ee
where $\p_I = \p/\p\Gamma^I$ are derivatives w.r.t. the coordinates $\Gamma^I\,(I=1,\dots,2N)$
of the phase space and
\be
\omega_0^{IJ}\,=\,\,\left(
\begin{array}{cc}
0 & 1_{N \times N} \\
-1_{N \times N} & 0
\end{array}
\right)
\ee
We can easily prove the Jacobi identity for the PB
\bn
\{A,\{B,C\}\}\;&+&\;\{B,\{C,A\}\}\;+\;\{C,\{A,B\}\}\,=\, \nonumber \\
&=&\omega_0^{IJ}\,\omega_0^{KL}\,\p_I A\,(\p_J \p_K B)\,\p_L C\;+\;\omega_0^{IJ}\,\omega_0^{KL}\,\p_I C\,\p_K A\,(\p_J \,\p_L B)\;+\;\ldots \nonumber \\
&=&0
\en
where we used that $\omega_0^T=-\omega_0$ besides the obvious fact that $\p_I\omega_0^{JK}=0$.  For the DB we can similarly write:
\be
\{A,B\}_{DB}\,=\,\p_I A\,\p_J B \,\omega^{IJ}
\ee
where
\bn
\omega^{IJ}\,=\,\omega_0^{IJ}\;-\;\omega_0^{IK}\,\Lambda_{KL}\,\omega_0^{LJ} \\
\Lambda_{KL}\,=\,\p_K \chi_{\alpha}\,\p_L \chi_{\beta}\,(C^{-1})^{\alpha\beta}
\en
For bosonic constraints $\chi_{\alpha}=0$ we have $\Lambda^T=-\Lambda$ and consequently:
\be
\omega^T\,=\,-\,\omega\;\;.
\ee
After a long calculation one can show from (9) and (10) the identity:
\be
\omega^{IJ}\,\p_J\,\omega^{KL}\,+\,\omega^{KJ}\,\p_J\,\omega^{LI}\,+\,\omega^{LJ}\,\p_J\,\omega^{IK}\,
\,=\,0 \;\;.
\ee

The Jacobi identity for DB (8) follows from (11) and (12):
\be
\{A,\{B,C\}_{DB}\}_{DB}\,+\,\ldots\,=\,0
\ee
As in \cite{dgl} we stress that in the quantization procedure DB will be replaced by commutators which identically satisfy the Jacobi identity.  Therefore it is physically mandatory that the Jacobi identity be obeyed at classical level.

\section{Dirac-like brackets}

Sometimes in order to keep relativistic covariance we write down the SCC in a redundant way.  This happens for instance in the case of the Brink-Schwarz \cite{bs} superparticle which is known to possess $8$ fermionic SCC while Majorana-Weyl spinors have 16 components in $D=10$.  Therefore when writing the constraints as a Majorana-Weyl spinor, eight components must be redundant, the same problem is present in the $D=10$ superstring \`a la Green and Schwarz.  Similarly, in many models for relativistic fractional spin particle in $2+1$ dimensions one assumes that the spin tri-vector is parallel to the momentum which might be imposed as  a tri-vector constraint $\chi_{\mu}=\epsilon_{\mu\nu\alpha}\,S^{\nu}\,P^{\alpha}=0$.  Clearly only two components of $\chi_{\mu}$ are independent and one component of $\chi_{\mu}$ is redundant, i.e., the constraints satisfy strongly the equation $P^{\mu}\chi_{\mu}=0$.  In general, redundant (or reducible) constraints satisfy reducibility conditions which can be written as
\be
Z^{a\beta}\,\chi_{\beta}\,=\,0
\ee
with $Z^{a\alpha}\;(a=1,\ldots,p;\alpha=1,\ldots,n)$ being specific functions of the phase-space coordinates.  There might be further stages of reducibility when the $Z^{a\alpha}$ themselves satisfy reducibility conditions on their own, like $\tilde{Z}_{\mu a}\,Z^{a\alpha}=0$, etc..  Concerning the Jacobi identity the number of stages of reducibility will prove to be immaterial since we only make use of (14).  From (14) we have $\{\chi_{\alpha},Z^{a\beta}\,\chi_{\beta}\}\,=\,0$ and consequently
\be
C_{\alpha\beta}\,Z^{a\beta}\,=\,-\{\chi_{\alpha},Z^{a\beta}\}\,\chi_{\beta}\;\;.
\ee
Therefore on the surface $\chi_{\beta}=0$ the quantities $Z^{\alpha\beta}$ become zero-modes of the matrix $C_{\alpha\beta}$ which imply that the DB defined in (4) can not be defined since $(C^{-1})^{\alpha\beta}$ does not exist.
Although $C_{\alpha\beta}$ can not be inverted, let us make an Ansatz for a new bracket:
\be
\{A,B\}^{\ast}\,=\,\{A,B\}\;-\;\{A,\chi_{\alpha}\}\,M^{\alpha\beta}\,\{\chi_{\beta},B\}
\ee
where the matrix $M^{\alpha\beta}$ is to be determined.  Therefore
\be
\{A,\chi_{\gamma}\}^{\ast}\,=\,\{A,\chi_{\gamma}\}\,-\,\{A,\chi_{\alpha}\}\,M^{\alpha\beta}\,C_{\beta\gamma}
\ee
Introducing a matrix $R^{\alpha}_{\gamma}$ associated with $M^{\alpha\beta}$
\be
M^{\alpha\beta}\,C_{\beta\gamma}\,\equiv\,\delta^{\alpha}_{\gamma}\,-\,R_{\gamma}^{\alpha}
\ee
we can write
\be
\{A,\chi_{\gamma}\}^{\ast}\,=\,\{A,\chi_{\alpha}\,R^{\alpha}_{\gamma}\}\,-\,\chi_{\alpha}\,\{A,R_{\gamma}^{\alpha}\}
\ee
Thus the new bracket will correctly eliminate the constraints, i.e., $\{A,\chi_{\gamma}\}^{\ast}=0$, if we assume $R^{\alpha}_{\gamma}=\lambda_{\gamma a}\,Z^{a\alpha}$ where $\lambda_{\gamma a}$ are in principle arbitrary.  In general the elimination of $\chi_{\alpha}$, different from the usual DB, will only occur on the surface $\chi_{\alpha}=0$.   Anyway since we are just looking for a self-consistent way to eliminate the SCC we only require that the r.h.s. of (19) vanishes on $\chi_{\alpha}=0$ surface.  Therefore all we need to consistently eliminate redundant SCC is a matrix $M^{\alpha\beta}$ such that
\be
M^{\alpha\beta}\,C_{\beta\gamma}\,=\,\delta^{\alpha}_{\gamma}\,-\,\lambda_{\gamma a}\,Z^{a\alpha}
\ee
the $\lambda_{\gamma a}$ are in principle arbritrary but consistency with the existence of zero modes (Eq. (15)) requires the $\lambda_{\gamma a}$ to satisfy on the $\chi_{\alpha}=0$ surface:
\be
Z^{a\beta}\,\lambda_{\beta b}\,Z^{b\delta}\,=\,Z^{a\delta}
\ee

Multiplying (21) by $\lambda_{\gamma a}$ we deduce on the
$\chi_{\alpha}=0$ surface

$$(\lambda\,Z)^2\,=\,\lambda\,Z$$

\noindent and so we are led to define the projection operators:
\bn
\lambda_{\alpha a}\,Z^{a\beta}\,&\equiv&\,(P_{\bot})^{\;\;\;\beta}_{\alpha} \nonumber \\
M^{\beta\delta}\,C_{\delta\alpha}\,=\,(1-P_{\bot})^{\;\;\;\beta}_{\alpha}\,&\equiv&\,(P_{\|})^{\;\;\;\beta}_{\alpha} \nonumber
\en
which satisfy
\bn
(P_{\|})^2\,&=&\,P_{\|} \nonumber \\
(P_{\bot})^2\,&=&\,P_{\bot} \nonumber \\
P_{\|}\,P_{\bot}\,&=&\,0\,=\,P_{\bot}\,P_{\|}
\en
on the $\chi_{\alpha}=0$ surface.  From the definitions we have
\be
P_{\bot}\,\chi\,=\,(\,1\,-\,P_{\|}\,)\,\chi\,=\,0\;\;.
\ee

Mathematically, our hypothesis (20) amounts to assume that the projection operator on the subspace of the SCC constraints $(P_{\|})^{\beta}_{\alpha}$ can be constructed from the matrix $C_{\alpha\beta}$ by multiplication: $(P_{\|})^{\beta}_{\alpha}=M^{\beta\gamma}C_{\gamma\alpha}$.  Clearly the matrix $M^{\alpha\beta}$ which satisfies (20) is not unique.  The change
\be
M^{\beta\gamma}\;\rightarrow\;M^{\beta\gamma}\,+\,(\,d^{\gamma}_a Z^{a\beta}\,-\,d^{\beta}_a Z^{a\gamma}\,)
\ee
for arbitrary $d_a^{\gamma}$ keeps (20) invariant.  Consequently the new brackets (17) are also apparently not unique.  However, the change (24) can be absorbed inside the Poisson brackets with the SCC in (17) and on the $\chi_{\alpha}=0$ surface the new brackets $\{A,B\}^{\ast}$ remain unchanged.

\section{Proof of Jacobi identity on $\chi_{\alpha}=0$ surface}

Clearly the new brackets $\{A,B\}^{\ast}$ defined in (16) and (20) can also be written in terms of a simpletic matrix just like the DB in (8) where we replace $(C^{-1})^{\alpha\beta}$ by $M^{\alpha\beta}$.  However the explicit proof of identity (12) relies heavily on the existence of $(C^{-1})^{\alpha\beta}$.  Thus it is not clear whether the new bracket will satisfy the Jacobi identity.
At this point we notice that there is a much simpler proof of the Jacobi identity for DB in \cite{ht} though it only holds on the $\chi_{\alpha}=0$ surface.  Since the new brackets $\{A,B\}^{\ast}$ eliminate the SCC only on the $\chi_{\alpha}=0$ surface it only makes sense to require the Jacobi identity on such surface.  Therefore we just have to adapt the proof of \cite{ht} replacing $(C^{-1})^{\alpha\beta}$ by $M^{\alpha\beta}$.  This is straightforward but we repeat it below for the reader's convenience.

The first step is to associate with any function in the phase space $F(q,p)$, a corresponding ``star" quantity $F^{\ast}(q,p)$:
\be
F^{\ast}\,=\,F\,-\,\{F,\chi_{\delta}\}\,M^{\delta\alpha}\,\chi_{\alpha}
\ee
From which it is obvious that
\be
\{F^{\ast},G\}\,=\,\{F,G\}^{\ast}\,+\,{\cal O}(\chi_{\alpha})
\ee
Replacing $G$ by $G^{\ast}$ in (26) and using that

$$
\{F,G^{\ast}\}^{\ast}\,=\,\{F,G\}^{\ast}+{\cal O}(\chi_{\alpha})\;\;.
$$

\noindent On the $\chi_{\alpha}=0$ surface  we have:
\be
\{F^{\ast},G^{\ast}\}\,=\,\{F,G\}^{\ast}
\ee

By successive use of (27) and (26) we derive respectively
\be
\{H,\{F,G\}^{\ast}\}^{\ast}\,=\,\{H,\{F^{\ast},G^{\ast}\}\}^{\ast}\,=\,
\{H^{\ast},\{F^{\ast},G^{\ast}\}\}
\ee
It is now clear that the Jacobi identity for the new brackets holds on the $\chi_{\alpha}=0$ surface as a direct consequence of the Jacobi identity for PB involving ``star" quantities:
\bn
\{H,\{F,G\}^{\ast}\}^{\ast}\,+\,\mbox{cyclic}\,=\,\{H,\{F,G\}\}\,+\,\mbox{cyclic}\,=\,0
\en
the proof is the same of \cite{ht} for DB since no particular use has been made of the existence of $(C^{-1})^{\alpha\beta}$.  We stress that (29) only holds on the $\chi_{\alpha}=0$ surface, which is not surprising, since the new brackets themselves only eliminate the constraints $\chi_{\alpha}$ on such surface.

Next we work out two particular examples in order to illustrate our general proof.

\section{Examples}

\subsection{Brink-Schwarz superparticle in $D=10$}

The action for the Brink-Schwarz massless superparticle in $D=10$ can be written as
\be
S\,=\,\int\,d\tau\,\left[\,p \cdot \dot{x}\,-\,i\,\theta\,\tilde{\pslash}\,\dot{\theta}\,-\,\frac{e\,p^2}{2}\,\right]
\ee
where $\theta_{\alpha}\;(\alpha=1,\ldots,16)$ is a Majorana-Weyl spinor in $D=10$ and $\tilde{\pslash}_{\alpha\beta}=p_{\mu}\tilde{\sigma}^{\mu}_{\alpha\beta}\;(\mu=0,1,\ldots,9)$.  The real and symmetric matrices $\tilde{\sigma}^{\mu}$ and $\sigma^{\mu}$ satisfy:
\bn
\left(\sigma^{\mu}{\tilde\sigma}^{\nu} +
\sigma^{\nu}\tilde\sigma^{\mu}\right)^{\alpha}_{\beta}&=&\left(
\tilde\sigma^{\mu}{\sigma}^{\nu} +
\tilde\sigma^{\nu}\sigma^{\mu}\right)^{\alpha}_{\beta}=2\, \eta^{\mu\nu}
\delta^{\alpha}_{\beta} \; \\
Tr(\tilde\sigma^{\mu}{\sigma}^{\nu}) &=& Tr(\sigma^{\mu}{\tilde\sigma}^{\nu})=
16\,\eta^{\mu\nu}
\en
and the special identity involving symmetrization of the indices\footnote{We use the notation of \cite{denis}.}:
\be
(\sigma^{\mu})^{(\alpha\beta}(\sigma_{\mu})^{\gamma\delta)}=0=
(\tilde\sigma^{\mu})_{(\alpha\beta}(\tilde\sigma_{\mu})_{\gamma\delta)}
\quad  ,
\ee
In total there are two bosonic constraints:
\bn
& &\phi_1 \quad : \quad p^2 \approx 0 \;\;, \\
& &\phi_2 \quad : \quad \Pi_e \approx 0 \;\; ,
\en
and sixteen fermionic ones $(\pi_{\alpha}=\p^R{\cal L}/\p\dot{\theta}^{\alpha})$
\be
d_\alpha = \Pi_{\alpha} + {\it i}(\pslash\theta)_{\alpha} \approx 0 \quad .
\ee

As shown in \cite{denis} the $16$ constraints $d_{\alpha}=0$
are equivalent to $8$  FCC $(\pslash\,d)_{\alpha}=0$ and $8$ SCC
\cite{bc}: 
\be 
\chi_{\alpha}\,=\,(\nslash\,d)_{\alpha}\,\approx\,0
\ee 
which are written however in a redundant way as a $16$
component spinors. 

The light-like vector $n^{\mu}$ can be chosen either as light cone constant vectors $n_{\pm}^{\mu}=(\pm1,0,\ldots,1)$ or (see \cite{denis}) as  the combinations below: 
\be 
n_{\pm}^{\mu} \; = \; x^2
p^{\mu} - [\, x\cdot p \pm ((x\cdot p)^2 - x^2p^2)^{1\over 2}\,]\,
x^{\mu} \quad \;\;. 
\ee

The second choice keeps Lorentz covariance but breaks supersymmetry explicitly.  We will concentrate on the second one henceforth, choosing in particular $n^{\mu}=n^{\mu}_+$.  On the surface $\chi_{\alpha}=0$ we have the PB:
\be
\{(\nslash\,d)^{\alpha},(\nslash\,d)^{\beta}\}\,=\,\frac{g}{(n \cdot p)}\,(\nslash)^{\alpha\beta} \,\equiv\,C^{\alpha\beta}
\ee
where\footnote{Notice that our definition of
$g$ differs from \cite{denis} by the factor $(x \cdot
p\,+\,f^{1\over2})$ which is missing in
\cite{denis}.},
\bn
g&=&4\,{\it i}\,(n\cdot p)\,[\,2\,n\cdot p\,-\,(x\cdot p\,+\,f^{1\over2})\,\Pi\cdot \theta\,] \\
f&=&(x\cdot p)^2\,-\,x^2\,p^2 
\en 
The constraints
$\chi_{\alpha}=(\nslash\,d)_{\alpha}$ satisfy the reducibility
condition 
\be
(\tilde{\nslash})_{\alpha\beta}\chi^{\beta}\,\equiv\,Z_{\alpha\beta}\chi^{\beta}=\,0
\ee leading to the zero modes \be
C^{\alpha\beta}\,Z_{\beta\gamma}\,=\,\frac{g}{2(n\cdot
p)}(\nslash\tilde{\nslash})^{\alpha}_{\;\;\;\gamma}\,=\,0 \ee
Relation (20) can be written as \be {g\over2(n\cdot
p)}\,M_{\gamma\beta}\,\nslash^{\beta\alpha}\,+\,\tilde{\nslash}_{\gamma\beta}\,\lambda^{\beta\alpha}\,=\,\delta^{\;\;\;\alpha}_{\gamma}
\ee On the other hand from the Clifford algebra (31) we have, for
an arbitrary vector $a_{\mu}$: \be \frac{\tilde{\nslash}
\,\aslash}{2(n\cdot
a)}\,+\,\frac{\tilde{\aslash}\,\nslash}{2(n\cdot a)}\,=\,1 \ee A
comparison naturally leads to the choice \bn
M_{\gamma\beta}&=&\frac{(n\cdot p)}{g}\,\frac{(\tilde{\aslash)}_{\gamma\beta}}{(n\cdot a)} \\
{\lambda}^{\gamma\beta}&=&\frac{(\aslash)^{\gamma\beta}}{2 (n\cdot a)}
\en
Any vector $a_{\mu}$ such that $n \cdot a \not= 0$ would be a possible choice.  We could choose $x_{\mu}$ or $p_{\mu}$ for instance.  The difference between these choices is proportional to $n_{\mu}$ that can always be added to $a_{\mu}$ which corresponds to add $Z_{\beta\gamma}$ to $M_{\beta\gamma}$ and as we remarked after (24) this will not change the brackets.  Following \cite{denis} we choose $a_{\mu}=p_{\mu}$ and assume $p \cdot n \not= 0$.  Therefore we have the new brackets
\be
\{A,B\}^{\ast}\,=\,\{A,B\}\,-\,\{A,(\nslash\, d)^{\alpha}\}\,\frac{(\tilde{\pslash})_{\alpha\beta}}{g}\{(\nslash\, d)^{\beta},B\}
\ee
From which we can reproduce the brackets of \cite{denis}, recalling that the definition of $g$ in \cite{denis} should be replaced by (40).  After this replacement we can show, after a long calculation that, by virtue of (33), the Jacobi identity, e.g.,
\bn
& &\{\theta^{\alpha},\{\theta^{\beta},\theta^{\gamma}\}^{\ast}\}^{\ast}\,+\,
\{\theta^{\beta},\{\theta^{\gamma},\theta^{\alpha}\}^{\ast}\}^{\ast}\,+\,
\{\theta^{\gamma},\{\theta^{\alpha},\theta^{\beta}\}^{\ast}\}^{\ast}\,= \nonumber \\
&=&\,-\,\frac{4{\it i}}{g^2}\,(n\cdot p)^2\,(x\cdot p\,+\,f^{1\over2})\,(\nslash\,\theta)_{\epsilon}\,
[\,(\sigma^{\mu})^{\epsilon\alpha}(\sigma_{\mu})^{\beta\gamma}\,+\,
(\sigma^{\mu})^{\epsilon\beta}(\sigma_{\mu})^{\gamma\alpha}\,+\,
(\sigma^{\mu})^{\epsilon\gamma}(\sigma_{\mu})^{\alpha\beta}\,] \nonumber \\
&=&0\;\;.
\en
We also have successfully checked explicitly other Jacobi identities involving $\{\pi_{\alpha},\{\theta_{\beta},\theta_{\gamma}\}^{\ast}\}^{\ast}$ and $\{\pi_{\alpha},\{\pi_{\beta},\theta_{\gamma}\}^{\ast}\}^{\ast}$.  In all these checks we have neglected terms proportional to $\chi_{\alpha}=(\nslash d)_{\alpha}$.  We certainly expect that all brackets given in \cite{denis} will satisfy the Jacobi identity on the $\nslash\,d=0$ surface.

The authors of \cite{dgl} have checked that some of the brackets calculated in \cite{denis} do not obey the Jacobi identity.  We do agree on that but the reason why they fail is because the definition of $g$ in \cite{denis} is incorrect and should be replaced by (40).  No extra changes are needed.  So we conclude that the Lorentz covariant but supersymmetry breaking constraints separation suggested in \cite{denis} is as self-consistent as the light-cone (supersymmetry preserving and Lorentz breaking) separation, at least at classical level.  As in \cite{denis} we claim that this indicates a competition between Lorentz and supersymmetry covariance in the massless superparticle.  Though Lorentz covariant, it should be emphasized that the algebra involved in the separation based on the vector $n_{\mu}$ given in (38) is more complicated than the light-cone one (see \cite{evans}).  In particular the gauge choice $\nslash \theta =0$ does not render the Lagrangian (30) quadratic as in the light-cone gauge $\Gamma^+ \theta =0$.  As far as we know there is no Lorentz covariant gauge choice that makes (30) quadratic without introducing extra degrees of freedom.  See \cite{berkovits} for a recent
Poincar\'e covariant quantization of the massless superparticle introducing $22$ extra fields.

\subsection{Fractional spin particle with spin and momentum parallel}

In terms of PB the Poincar\'e algebra in $2+1$ dimensions is given by
\bn
\{P_{\mu},P_{\nu}\}&=&0 \nonumber \\
\{J_{\mu},P_{\nu}\}&=&\epsilon_{\mu\nu\alpha}\,P^{\alpha} \nonumber \\
\{J_{\mu},J_{\nu}\}&=&\epsilon_{\mu\nu\alpha}\,J^{\alpha}\;\;,
\en
where $\mu,\nu,\alpha=0,1,2$ and $J_{\mu}$ is the dual angular momentum, i.e., $J_{\mu}={1\over2}\epsilon_{\mu\nu\alpha}\,J^{\nu\alpha}$.  Introducing canonical, conjugated variables satisfying,
\bn
\{x_{\mu},x_{\nu}\}&=&0 \nonumber \\
\{x_{\mu},p_{\nu}\}&=&\eta_{\mu\nu} \nonumber \\
\{p_{\mu},p_{\nu}\}&=&0\;\;,
\en
and some extra variables $(q_{i},\pi_{j})$ to describe spin $S_{\mu}(q_{i},\pi_{j})$, we can realize the algebra (53) by means of
\bn
P_{\mu}\,&=&\,p_{\mu} \nonumber \\
J_{\mu}\,&=&\,\epsilon_{\mu\nu\alpha}\,x^{\nu}\,p^{\alpha}\,+\,S_{\mu}(q_{i},\pi_{j})\;\;,
\en
where $S_{\mu}\,(q_{i},\pi_{j})$ must be such that the angular momentum algebra is satisfied,
\be
\{ S_\mu,S_\nu\}=\varepsilon_{\mu\nu\lambda}S^\lambda\ .
\ee
and we assume that
\be
\{S_{\mu},n_{\alpha}\}\,=\,0\,=\,\{S_{\mu},p_{\alpha}\} \;\;.
\ee

The algebra (53) possess two Casimir invariants which fix the mass and the helicity $(\alpha\,m)$ of the particle:
\bn
P^2\,-\,m^2\,&=&\,0\ \;\;, \\
J \cdot P\,+\,\alpha\,m\,&=&\,S \cdot P\,+\,\alpha\,m\,=\,0\;\;,
\en
where $\alpha$ may be any real number.  There is no need for having only integer or half-integer spin in $2+1$ dimensions.

In $2+1$ dimensions, particles with higher spin do not have more degrees of freedom than lower spin particles.  There is always only one polarization state and the spin does not represent extra degrees of freedom.  Since the component of the spin parallel to the momentum is fixed by the Casimir invariant in (56) it is natural to impose that no extra components exist and so $S_{\mu}$ and $p_{\mu}$ are parallel.  This is indeed a common, though not necessary, feature of many models for fractional spin particles in the literature \cite{dd,cnp,cfm,dd2}.  In order to impose that $S_{\mu}$ and $p_{\mu}$ be parallel in a model independent way, one must use the tri-vector constraint \cite{dd2}:
\be
\chi_{\mu}\,=\,\epsilon_{\mu\nu\alpha}\,S^{\nu}\,p^{\alpha}\,\approx\,0
\ee
which are second-class:
\be
C_{\alpha\beta}\,=\,\{\chi_{\alpha},\chi_{\beta}\}\,=\,\epsilon_{\alpha\beta\gamma}p^{\gamma}\,(S \cdot p)
\ee
and since: $p^{\mu}\,\chi_{\mu}=0$ identically, the matrix $C_{\alpha\beta}$ possess the zero mode $Z_{\mu}=p_{\mu}$, i.e., $C_{\alpha\beta}\,Z^{\beta}=0$.  Therefore $(C^{-1})_{\alpha\beta}$ does not exist and we have to introduce new brackets:

$$\{A,B\}^{\ast}\,=\,\{A,B\}\;-\;\{A,\chi_{\alpha}\}\,\left(\frac{\epsilon^{\alpha\beta\gamma}\,K_{\gamma}}{(K \cdot p)\,C}\right)\,\{\chi_{\beta},B\}$$

\noindent as in (16).  The vector $K_{\alpha}$ is arbitrary but such that $K \cdot p \not=0$, and $C={\it l}p^2$ where, from (59),
\be
{\it l}\,=\,\frac{S \cdot p}{p^2}\;\;.
\ee

The brackets (51) will be replaced by:
\bn
\{x_{\mu},x_{\nu}\}^{\ast}&=&-\,\frac{\alpha\,m}{p^2}\,\epsilon_{\mu\nu\lambda}\,p^{\lambda} \\
\{x_{\mu},p_{\nu}\}^{\ast}&=&\eta_{\mu\nu} \\
\{p_{\mu},p_{\nu}\}^{\ast}&=&0
\en
We have checked for example the Jacobi identity between the coordinates $x^{\mu}$.  From (60) and (61) we have
\be
\{x_{\mu},\{x_{\nu},x_{\lambda}\}^{\ast}\}^{\ast}\,=\,-\,\frac{\alpha\,m}{p^2}\,\left[\,\epsilon_{\mu\nu\lambda}\,-\,3\,\frac{p_{\nu}\epsilon_{\lambda\mu\alpha}\,p^{\alpha}}{p^2} \right]
\ee
and,
\be
\{x_{\mu},\{x_{\nu},x_{\lambda}\}^{\ast}\}^{\ast}\,+\,\{x_{\nu},\{x_{\lambda},x_{\mu}\}^{\ast}\}^{\ast}\,+\,\{x_{\lambda},\{x_{\mu},x_{\nu}\}^{\ast}\}^{\ast}\,=\,0\;\;,
\ee
where we have used the identity:
\be
P^2\,\epsilon_{\nu\lambda\mu}\,=\,P_{\nu}\,\epsilon_{\lambda\mu\alpha}\,P^{\alpha}\,+\,P_{\lambda}\,\epsilon_{\mu\nu\alpha}\,P^{\alpha}\,+\,P_{\mu}\,\epsilon_{\nu\lambda\alpha}\,P^{\alpha}
\ee

From our general proof we believe, of course, that all new brackets will satisfy the Jacobi identity.

\section{Conclusion}

For redundant constraints the Dirac brackets can not be defined
since the matrix $C_{\alpha\beta}=\{\chi_\alpha,\chi_\beta\}$ has
no inverse.  However, we have shown that new brackets can be
introduced based on the hypothesis that a covariant projection
operator on the space of redundant second-class constraints can be
constructed from $C_{\alpha\beta}$ by multiplication by some
matrix
$M^{\alpha\beta}:\;\;(P_{\|})^{\alpha}_{\gamma}=M^{\alpha\beta}
\,C_{\beta\gamma}$ and $P_{\|}\,\chi=\chi$.  We have presented a
general proof that on the surface of the second-class constraints
$\chi_{\alpha}=0$ the Jacobi identity for those new brackets will
always be satisfied.  In particular, we have checked it explicitly
for the Brink-Schwarz massless superparticle in $D=10$ and
fractional spin particles in $2+1$ dimensions with spin and
momentum parallel.

\section{Acknowledgments}

E.M.C.A. is finantially supported by Funda\c{c}\~ao de
Amparo \`a Pesquisa do Estado de S\~ao Paulo (FAPESP). This work  is also partially supported by 
Conselho Nacional de Desenvolvimento Cient\'{\i}fico e Tecnol\'ogico (CNPq).  FAPESP and CNPq are brazilian research agencies.


\newpage

\begin{thebibliography}{99}
\bibitem{dirac}  P.A.M. Dirac, ``Lectures on Quantum Mechanics", Academic Press, Yeshiva University, New York, 1967.

\bibitem{dgl}   A. A. Deriglazov, A. N. Galajinski and S. L. Lyakhovich, Nucl. Phys. B 473 (1996) 245.

\bibitem{bs}  L. Brink and J. H. Schwarz, Phys. Lett. B 128 (1983) 397.

\bibitem{ht}   M. Henneaux and C. Teitelboim, ``Quantization of Gauge Systems", Princeton University Press, Princeton, New Jersey, 1982..

\bibitem{denis}  D. Dalmazi, Phys. Lett. B 328 (1994) 43.

\bibitem{bht}  L. Brink, M. Henneaux and C. Teitelboim, Nucl. Phys. B 293 (1987) 505.

\bibitem{bc}  I. Bengtsson and M. Cederwall, G\"oteborg preprint 84-21 (1984), unpublished.

\bibitem{evans}  J. M. Evans, Class. Quantum Grav. 7 (1990) 699.

\bibitem{berkovits}   N. Berkovits, ``Covariant quantization of the superparticle using pure spinors", hep-th/0105050.

\bibitem{dd}  D. Dalmazi and A. de Souza Dutra, Phys. Lett. B 414 (1997) 315.

\bibitem{cnp}   Chi-hong Chou, V. P. Nair and Alexios P. Polychronakos,
Phys. Lett. B 304 (1993) 105.

\bibitem{cfm}  M. Chaichian, R. G. Felipe and D. L. Martinez, Phys. Rev. Lett. 71 (1993) 3045.

\bibitem{dd2}  D. Dalmazi and A. de Souza Dutra, Phys. Lett. B 343 (1995) 225.

\bibitem{plyushchay}  M.S. Plyushchay, Int. J. of Mod. Phys. {\bf A6} (1992) 7045.

\bibitem{jn}  R. Jackiw and V.P. Nair, Phys. Rev. {\bf D43} (1991) 1933.

\bibitem{binegar}  B. Binegar, J. Math. Phys. {\bf 23} (1982) 1511.
\end{thebibliography}
\end{document}